\documentclass[acus]{JAC2001}

\usepackage{latexsym}
\usepackage{hyperref}

\newcommand{\etal}{{\em et al.}}

\newcommand{\gev}{\hbox{ GeV}}

\newcommand{\tevcc}{\hbox{ TeV}\!/\!c^2}

\newcommand{\mm}{\hbox{ mm}}

\newcommand{\m}{\hbox{ m}}
\newcommand{\lum}{\hbox{ cm}^{-2}\hbox{ s}^{-1}}

\def\url#1{\mbox{\href{#1}{\sf #1}}}

\newcommand{\hepex}[1]{hep-ex/#1}
\newcommand{\hepph}[1]{hep-ph/#1}

\def\prll#1#2#3{\frenchspacing{\it Phys. Rev. Lett. }{\bf #1}, #2 (#3)}
\def\pr#1#2#3{\frenchspacing{\it Phys. Rev. D}{\bf #1}, #2 (19#3)}
\def\prM#1#2#3{\frenchspacing{\it Phys. Rev. D}{\bf #1}, #2 (#3)}

\def\pl#1#2#3{\frenchspacing{\it Phys. Lett. }{\bf #1}, #2 (19#3)}
\def\np#1#2#3{\frenchspacing{\it Nucl. Phys. }{\bf #1}, #2 (19#3)}
\def\npM#1#2#3{\frenchspacing{\it Nucl. Phys. }{\bf #1}, #2 (#3)}

\def\app#1#2#3{\frenchspacing{\it Acta Phys. Polon. B}{\bf #1}, #2 (19#3)}

\def\phystoday#1#2#3#4{\frenchspacing{\it Phys. Today\/ }{\bf #1}, 
(#3) #2 
(\ifcase#3\or January\or 
         February\or March\or April\or May\or June\or July\or August\or 
         September\or October\or November\or December\fi\ 19#4)}

\begin{document}
\title{PARTICLE PHYSICS---FUTURE DIRECTIONS}

\author{Chris Quigg\thanks{quigg@fnal.gov}, Fermi National Accelerator 
Laboratory, Batavia, IL 60510, USA}

\maketitle

\begin{abstract}
Wonderful opportunities await particle physics over the next decade,
with the coming of the Large Hadron Collider at CERN to explore the
1-TeV scale (extending efforts at LEP and the Tevatron to unravel
the nature of electroweak symmetry breaking) and many initiatives to
develop our understanding of the problem of identity: what makes a
neutrino a neutrino and a top quark a top quark.  Here I have in mind
the work of the $B$ factories and the Tevatron collider on \textsf{CP} violation
and the weak interactions of the $b$ quark; the wonderfully sensitive
experiments at Brookhaven, CERN, Fermilab, and Frascati on \textsf{CP}
violation and rare decays of kaons; the prospect of definitive
accelerator experiments on neutrino oscillations and the nature of the
neutrinos; and a host of new experiments on the sensitivity frontier. 
We might even learn to read experiment for clues about the
dimensionality of spacetime.  If we are inventive enough, we may be
able to follow this rich menu with the physics opportunities offered
by a linear collider and a (muon storage ring) neutrino factory.  I
expect a remarkable flowering of experimental particle physics, and of
theoretical physics that engages with experiment.  I  describe 
some of the
great questions before us and the challenges of providing the
instruments that will be needed to define them more fully
and--eventually--to answer them.
\end{abstract}

\section{PARTICLE PHYSICS \\ AT THE MILLENNIUM}

The physics curriculum in the 1898--99 University of 
Chicago catalogue begins with a very Victorian preface~\cite{sbt}:
\begin{quote}
    ``While it is never safe to affirm that the future of the Physical
    Sciences has no marvels in store even more astonishing than those of
    the past, it seems probable that most of the grand underlying
    principles have been firmly established and that further advances are
    to be sought chiefly in the rigorous application of these principles
    to all the phenomena which come under our notice \ldots .  An eminent
    physicist has remarked that the future truths of Physical Science are
    to be looked for in the sixth place of decimals.''
\end{quote}
As the ink was drying on these earnest words, R\"{o}ntgen discovered x
rays and published the epoch-making radiograph of his wife's hand,
Becquerel and the Curies explored radioactivity, Thomson discovered
the electron and showed that the ``uncuttable'' atom had parts, and
Planck noted that anomalies in the \textit{first} place of the
decimals required a wholesale revision of the laws of Nature.

We have the benefit of a century of additional experience and insight,
but we are not nearly so confident that we have uncovered ``most of
the grand underlying principles.''  Indeed, while we celebrate the
insights codified in the \textit{standard model of particle physics}
and look forward to resolving its puzzles, we are increasingly
conscious of how little of the physical universe we have experienced. 
Future truths are still to be found in precision measurements, but the
century we are leaving has repeatedly shown that Nature's marvels are
not limited by our imagination.  Exploration can yield surprises that
completely change what we think about---and how we think.

We base our understanding of physical phenomena on the identification
of a few constituents that seem elementary at the current limits of
resolution of about $10^{-18}\m$, and a few fundamental forces.  The
constituents are the pointlike quarks \{$(u, d)_{L}$, $(c, s)_{L}$, $(t,
b)_{L}$\} and leptons \{$(\nu_{e}, e)_{L}$, $(\nu_{\mu}, \mu)_{L}$,
$(\nu_{\tau}, \tau)_{L}$\}, with strong, weak, and electromagnetic
interactions specified by $SU(3)_{c}\otimes SU(2)_{L}\otimes U(1)_{Y}$
gauge symmetries.

The electroweak theory is founded on the weak-isospin symmetry 
embodied in the quark and lepton doublets
and weak-hypercharge phase symmetry, plus the idealization that 
neutrinos are massless.\footnote{For surveys of the electroweak
theory, with references, see Ref.\
\cite{Quigg:1999xg}.} In its simplest form, with the
electroweak gauge symmetry broken by the Higgs mechanism, the
$SU(2)_{L}\otimes U(1)_{Y}$ theory has scored many qualitative
successes: the prediction of neutral-current interactions, the
necessity of charm, the prediction of the existence and properties of
the weak bosons $W^{\pm}$ and $Z^{0}$.  Over the past ten years, in
great measure due to the beautiful experiments carried out at the $Z$
factories at CERN and SLAC, precision measurements have tested the
electroweak theory as a quantum field theory, at the one-per-mille
level~\cite{Sirlin:1999zc,Swartz:1999xv,Charlton:2001am}.
Last year, our colleagues working at LEP made a heroic push to discover 
the Higgs boson 2000~\cite{unknown:2001xw}.  The search will intensify 
again in a few years at the Tevatron and the Large Hadron Collider.

The quark model of hadron structure and the parton model of 
hard-scattering processes have such pervasive influence on the way we 
conceptualize particle physics that quantum chromodynamics, the theory 
of strong interactions that underlies both, sometimes is taken for granted.  
QCD is a remarkably simple, successful~\cite{Quigg:1999di}, and rich 
theory of the strong interactions~\cite{Wilczek:1999id}.  The 
perturbative regime of QCD exists, thanks to the crucial property of 
asymptotic freedom, and describes many phenomena in quantitative 
detail.  The strong-coupling regime controls hadron structure and 
gives us our best information about quark masses.  Unfamiliar
r\'{e}gimes of high density and high temperature contain riches we
have only begun to explore.

\section{What we need to know}
This concise statement of the standard model invites us to consider 
the agenda of particle physics today under a few broad rubrics.  

\vspace*{3pt}\noindent\textit{Elementarity.} Are the quarks and leptons structureless, or
will we find that they are composite particles with internal
structures that help us understand the properties of the individual
quarks and leptons?

\vspace*{3pt}\noindent\textit{Symmetry.} One of the most powerful lessons of the modern
synthesis of particle physics is that (local) symmetries prescribe
interactions.  Our investigation of symmetry must address the question
of which gauge symmetries exist (and, eventually, why).  We have
learned to seek symmetry in the laws of Nature, not (necessarily) in
the consequences of those laws.  Accordingly, we must understand how
the symmetries are hidden from us in the world we inhabit.  For the
moment, the most urgent problem in particle physics is to complete our
understanding of electroweak symmetry breaking by exploring the 1-TeV
scale.  This is the business of the experiments at LEP2, the Tevatron
Collider, and the Large Hadron Collider.

\vspace*{3pt}\noindent\textit{Unity.} In the sense of developing explanations that apply not
to one individual phenomenon in isolation, but to many phenomena in
common, unity is central to all of physics, and indeed to all of
science.  At this moment in particle physics, our
quest for unity takes several forms.

First, we have the fascinating possibility of gauge coupling
unification, the idea that all the interactions we encounter have a
common origin and thus a common strength at suitably high energy.

Second, there is the imperative of anomaly freedom in the electroweak
theory, which urges us to treat quarks and leptons together, not as
completely independent species.  Both of these ideas are embodied, of
course, in unified theories of the strong, weak, and electromagnetic
interactions, which imply the existence of still other forces---to
complete the grander gauge group of the unified theory---including
interactions that change quarks into leptons.

The third aspect of unity is the idea that the traditional distinction
between force particles and constituents might give way to a unified
understanding of all the particles.  The gluons of QCD carry color
charge, so we can imagine quarkless hadronic matter in the form of
glueballs.  Beyond that breaking down of the wall between messengers
and constituents, supersymmetry relates fermions and bosons.

Finally, we desire a reconciliation between the pervasive outsider,
gravity, and the forces that prevail in the quantum world of our
everyday laboratory experience.

\vspace*{3pt}\noindent\textit{Identity.} We do not understand the physics that sets quark
masses and mixings.  Although we are testing the idea that the phase
in the quark-mixing matrix lies behind the observed \textsf{CP}
violation, we do not know what determines that phase.  The
accumulating evidence for neutrino oscillations presents us with a new
embodiment of these puzzles in the lepton sector.  At bottom, the
question of identity is very simple to state: What makes an electron
and electron, and a top quark a top quark?

\vspace*{3pt}\noindent\textit{Topography.} ``What is the
dimensionality of spacetime?''  tests our preconceptions and unspoken
assumptions.  It is given immediacy by recent theoretical work.  For
its internal consistency, string theory requires an additional six or
seven space dimensions, beyond the $3+1$ dimensions of everyday
experience.  Until recently it has been presumed that the extra
dimensions must be compactified on the Planck scale, with a
stupendously small compactification radius $R \simeq
M_{\mathrm{Planck}}^{-1} = 1.6 \times 10^{-35}\m$.
Part of the vision of string theory is that what goes on in even such
tiny curled-up dimensions does affect the everyday world: excitations
of the Calabi--Yau manifolds determine the fermion
spectrum.\footnote{For a gentle introduction to the goals of
string theory, see Ref.\ \cite{beegee}.}

We have recognized recently that Planck-scale compactification is 
not---according to what we can establish---obligatory, and that current
experiment and observation admit the possibility of dimensions not 
navigated by the strong, weak, and electromagnetic interactions that 
are almost palpably large.  A whole range of new experiments will help 
us explore the fabric of space and time, in ways we didn't expect just 
a few years ago~\cite{EDsearch}.
\section{A DECADE OF DISCOVERY AHEAD}
Over the next decade, we may look forward to an avalanche of 
experimental results that have the potential to change our view of the
fundamental particles and their interactions in very dramatic ways.  A
special preoccupation for me is the search and study of the Higgs
boson; this is really shorthand for a thorough exploration of the
1-TeV scale, which will elucidate the mechanism of electroweak
symmetry breaking.  We can also expect wonderful progress in flavor
physics: the detailed study of \textsf{CP} violation in the $B$
system, dramatically increased sensitivity in the exploration of rare
decays of $K$ and $D$ mesons, and pinning down the nature of neutrino
oscillations.  And maybe we will at last see a \textsf{CP}-violating
permanent electric dipole moment of the neutron.  Run II of the
Tevatron will give us our first opportunity to use the top quark as a
tool, and not only as an object of desire.  Although the
interpretation of heavy-ion collisions at RHIC and the LHC promises to
be challenging, the heavy-ion colliders offer a real chance to
discover new phases of matter and enrich our understanding of QCD. On
many fronts, we are taking dramatic steps in energy and sensitivity
that will help us explore: extra dimensions, new dynamics,
supersymmetry, and new kinds of forces and constituents might show
themselves.  (I'm conflicted about whether I'd like to see them all at
once, or in easy-to-understand installments!)

Experiments that use natural sources also hold great promise for the
decade ahead.  We suspect that the detection of proton decay is only a
few orders of magnitude away in sensitivity.  Astronomical
observations should tell us what kinds of matter and energy make up
the universe.  Indeed, the whole complex of experiments and 
observations we call astro/cosmo/particle physics should enjoy a 
golden age.

Now, the decade of discovery won't happen automatically.  Many of our 
goals are difficult, and timely success is in doubt for many 
experiments.  We must push hard to prepare the instruments, and get to 
the answers.

The glorious future of new machines and new experiments that lies beyond 
the established program also won't happen by itself.  We have, I 
think, come to the collective realization that we must do more to 
prepare alternative futures by creating a rich and organic program of 
accelerator research.  We're also challenged by our success: the scope of our 
science has grown, but funding has not.  Within our own extended 
family and beyond, we must do more to convey the urgency and 
importance of the new scientific opportunities, and fashion a program 
that we can carry out that includes the right measure of scale 
diversity to ensure a healthy intellectual ecosystem.

\section{THE ORIGINS OF MASS}

A key aspect of the problem of identity is the origin of mass.  In fact,
we know the challenge of explaining many different kinds of mass.  The
masses of the hadrons are (in principle, and with increasing precision
in practice) understood from QCD in terms of the energy stored to
confine a color-singlet configuration of quarks in a small
volume~\cite{Wilczek:be,Aoki:1999yr}.  This is a remarkable achievement.  We also
have an excellent understanding of the masses of the electroweak gauge
bosons $W^{\pm}$ and $Z^{0}$ as consequences of electroweak symmetry
breaking, in terms of a single weak mixing parameter,
$\sin^{2}\theta_{W}$.\footnote{Although for the moment we take the
weak mixing parameter from experiment, we understand how it arises in
a unified theory.  Indeed, in a unified theory we can hope to
understand the parameter $\Lambda_{\mathrm{QCD}}$ that sets the scale
of the hadron masses.} At tree level in the electroweak theory, we
have
\begin{eqnarray*}
    M_{W}^{2} & = & g^{2}v^{2}/2 = 
    \pi\alpha/G_{F}\sqrt{2}\sin^{2}\theta_{W} ,
    \label{eq:Wmass}  \\
    M_{Z}^{2} & = & M_{W}^{2}/\cos^{2}\theta_{W} ,
    \label{eq:Zmass}  
\end{eqnarray*}
with the electroweak scale $v = (G_{F}\sqrt{2})^{-\frac{1}{2}}
\approx 246\gev$.

When we get to the question of quark and
(charged) lepton masses, however, our understanding is considerably
more primitive.  For each of these, we require not just the scale of
electroweak symmetry breaking, but a distinct and apparently arbitrary
Higgs-fermion-antifermion Yukawa coupling to reproduce the fermion
mass.  In the electroweak theory, the value of each quark or charged-lepton
mass is set by a new, unknown, Yukawa coupling.  Taking the electron
as a prototype, we define the left-handed doublet and right-handed
singlet
\begin{displaymath}
    \mathsf{L} = \left( 
    \begin{array}{c}
        \nu_{e}  \\
        e
    \end{array}
    \right)_{L} \; , \qquad \mathsf{R} \equiv e_{R}.
    \label{eq:elec}
\end{displaymath}
Then the Yukawa term in the electroweak Lagrangian is
\begin{displaymath}
    \mathcal{L}_{\mathrm{Yukawa}}^{(e)} = - 
    \zeta_{e}[\bar{\mathsf{R}}(\varphi^{\dagger}\mathsf{L}) + 
    (\bar{\mathsf{L}}\varphi)\mathsf{R}] \; ,
    \label{eq:eYuk}
\end{displaymath}
where $\varphi$ is the Higgs field, so that the electron mass is
$m_{e} = \zeta_{e}v/\sqrt{2}$.  For neutrinos, which may be their own
antiparticles, there are still more possibilities for new physics to
enter.  Inasmuch as we do not know how to calculate the fermion Yukawa
couplings $\zeta_{f}$, I believe that \textit{we should consider the
sources of all fermion masses as physics beyond the standard model.}

The values of the Yukawa couplings are vastly different for
different fermions: for the top quark, $\zeta_{t} \approx 1$, for the
electron $\zeta_{e} \approx 3\times 10^{-6}$, and if the neutrinos
have Dirac masses, presumably $\zeta_{\nu} \approx
10^{-10}$.\footnote{I am quoting the values of the Yukawa couplings at
a low scale typical of the masses themselves.} What accounts for the
range and values of the Yukawa couplings?  Our best hope until now has
been the suggestion from unified theories that the pattern of fermion
masses simplifies on high scales.  The classic intriguing prediction
of the $SU(5)$ unified theory involves the masses of the $b$ quark and
the $\tau$ lepton, which are degenerate at the unification point for a
simple pattern of spontaneous symmetry breaking.  The different
running of the quark and lepton masses to low scales then leads to the
prediction $m_{b} \approx 3 m_{\tau}$, in suggestive agreement with
what we know from experiment~\cite{Buras:1977yy}.

``Large'' extra dimensions present us with new ways to think about the
exponential range of Yukawa couplings.  If the standard-model brane
has a small thickness, the wave packets representing different fermion
species might have different locations within the extra
dimension~\cite{Arkani-Hamed:1999dc,Mirabelli:1999ks}.  On this
picture, the Yukawa couplings measure the overlap in the extra
dimensions of the left-handed and right-handed fermion wave packets and the
Higgs field, presumed pervasive.  Exponentially large differences
might then arise from small offsets in the new coordinate(s).  True or
not, it is a mind-expanding way to look at an important problem.

\section{GRAVITY AND EXTRA DIMENSIONS}
It is entirely natural to neglect gravity in most particle-physics 
applications, because the coupling of a graviton $\mathcal{G}$ to a particle is 
tiny, generically of order $(E/M_{\mathrm{Planck}})$ where $E$ is a 
typical energy scale of the problem.  Thus, for example, we expect 
the branching fraction $B(K \rightarrow \pi\mathcal{G}) \sim 
(M_{K}/M_{\mathrm{Planck}})^{2} \sim 10^{-38}$.  And yet we cannot 
put gravity entirely out of our minds, even if we restrict our 
attention to standard-model interactions at attainable energies.

The great gap between the electroweak scale of about $10^{3}\gev$ and the 
Planck scale of about $10^{19}\gev$
gives rise to the hierarchy problem of the  electroweak theory 
\cite{hier}: how to protect the Higgs-boson mass from quantum 
corrections that explore energies up to $M_{\mathrm{Planck}}$.
The conventional approach to the hierarchy problem has been to ask why
the electroweak scale (and the mass of the Higgs boson) is so much
smaller than the Planck scale.  Framing the issue this way leads us to 
change the electroweak theory to include supersymmetry, or 
technicolor, or some other extension. Over the past few years, we have begun
instead to ask why gravity is so weak.  This question motivates us to
consider \textit{changing gravity} to understand why the Planck scale
is so large \cite{EDbiblio}.  Now,
elegant experiments that study details of Casimir and van der Waals
forces tell us that gravitation closely follows the Newtonian force
law down to distances on the order of $0.3\mm$~\cite{Hoyle:2000cv}, 
which corresponds to an energy scale of only about $10^{-12}\gev$! 
At shorter distances (higher energies), the constraints on deviations from Newton's
inverse-square force law deteriorate rapidly, so nothing prevents us
from considering changes to gravity even on a small but macroscopic
scale.

One way to change the force law is to imagine that gravity can
propagate into extra dimensions.  To respect the stronger constraints
on the behavior of the standard-model interactions, we suppose that
the $SU(3)_{c}\otimes SU(2)_{L}\otimes U(1)_{Y}$ gauge fields, plus
needed extensions, reside on $3+1$-dimensional branes, not in the
extra dimensions.

How does this hypothesis change the picture?  The dimensional 
analysis (Gauss's law, if you like) that relates Newton's constant to 
the Planck scale changes.  If gravity propagates in $n$ extra 
dimensions with radius $R$, then
\begin{displaymath}
    G_{\mathrm{Newton}} \sim M_{\mathrm{Planck}}^{-2} \sim M^{\star\,-n-2}R^{-n}\; ,
    \label{eq:gauss}
\end{displaymath}
where $M^{\star}$ is gravity's true scale.  Notice that if we boldly 
take $M^{\star}$ to be as small as $1\tevcc$, then the radius of the extra 
dimensions is required to be smaller than about $1\mm$, for $n \ge 
2$.  If we use the four-dimensional force law to extrapolate the 
strength of gravity from low energies to high, we find that gravity 
becomes as strong as the other forces on the Planck scale.  If the force law 
changes at an energy $1/R$, as the large-extra-dimensions scenario 
suggests, then the forces are unified at lower energy $M^{\star}$.
What we know as the Planck scale is then a mirage that results 
from a false extrapolation: treating gravity as four-dimensional down 
to arbitrarily small distances, when in fact---or at least in this 
particular fiction---gravity propagates in $3+n$ spatial dimensions.  
The Planck mass is an artifact, given by $M_{\mathrm{Planck}} = 
M^{\star}(M^{\star}R)^{n/2}$.   If the true scale of gravity were close 
to $M_{H}$, the hierarchy problem would recede.

\section{$\nu$ OSCILLATION NEWS}
The science that grew into particle physics began with found
beams---the emanations from naturally occurring radioactive substance
and the cosmic rays---and found beams still provide us with important
windows on the universe.  One of the great scientific detective
stories of the recent past is the developing case for neutrino
oscillations: the evidence that neutrinos produced as one flavor
($\nu_{e}$, $\nu_{\mu}$, or $\nu_{\tau}$) actually morph into other
flavors.  Long known as a theoretical possibility, \textit{neutrino
oscillation} is now all but established by the Super-Kamiokande 
experiment's observation of an up-down asymmetry in the flux of muon 
neutrinos produced by the interaction of cosmic rays in the 
atmosphere~\cite{Toshito:2001dk}.  By far the most graceful interpretation 
is that $\nu_{\mu}$ produced on the other side of the Earth oscillate 
during flight in significant numbers into $\nu_{\tau}$.

Just as PAC2001 convened, the Sudbury Neutrino Observatory added an 
important new element to our understanding of the longstanding puzzle 
of the solar neutrino deficit~\cite{Ahmad:2001an}.  SNO reports an 
impressively precise measurement of the solar neutrino charged-current 
cross section on the heavy-water target that serves as their 
water-Cherenkov detector.  The measured rate implies a $\nu_{e}$ flux 
\begin{displaymath}
	{\phi^{\mathrm{CC}}_{\mathrm{SNO}}(\nu_{e}) = 1.75 \pm 0.07 ^{+0.12}_{-0.11} 
 \pm 0.05 \times 10^{6}\lum}\; ,
\end{displaymath}
where the uncertainties are statistical, systematic, and theoretical.  
They have also measured the solar neutrino elastic ($\nu_{x} e$) cross 
section with limited precision, and extracted from it the flux of 
solar neutrinos of all active flavors,
\begin{displaymath}
	\phi^{\mathrm{ES}}_{\mathrm{SNO}}(\nu_{x}) = 2.39 \pm 0.34 ^{+0.16}_{-0.14} 
 \times 10^{6}\lum \; .
\end{displaymath}
The SNO experimenters are in the right place at the right time, because the 
Super-K experiment has already given a very precise measurement of 
the solar neutrino flux from elastic ($\nu_{x} e$) scattering~\cite{Fukuda:2001nj},
\begin{displaymath}
	{\phi^{\mathrm{ES}}_{\mathrm{SK}}(\nu_{x}) = 
	2.32 \pm 0.03 ^{+0.08}_{-0.07} 
 \times 10^{6}\lum}\; .
\end{displaymath}
The difference between the flux of active neutrinos and the flux of 
electron neutrinos,
\begin{displaymath}
	{\phi^{\mathrm{ES}}_{\mathrm{SK}}(\nu_{x}) -
	\phi^{\mathrm{CC}}_{\mathrm{SNO}}(\nu_{e}) = 0.57 \pm 0.17\times
	10^{6}\lum}\; ,
\end{displaymath}
demonstrates at $3.3\sigma$ that active neutrinos other than $\nu_{e}$, namely
$\nu_{\mu}$ and $\nu_{\tau}$, arrive at Earth.  Since the nuclear 
processes that power the Sun yield only $\nu_{e}$, this new result 
rules in favor of neutrino oscillations as the explanation for the 
solar neutrino puzzle.

\section{THE QUEST FOR NEW TOOLS}
Although theoretical speculation and synthesis is valuable and
necessary, we cannot advance without new observations.  The
experimental clues needed to answer today's central questions can come
from experiments at high-energy accelerators, experiments at
low-energy accelerators and nuclear reactors, experiments with found
beams, and deductions from astrophysical measurements.  Past 
experience, our intuition, and the current state of particle theory 
all point to an indispensable role for accelerator experiments.

The opportunities for accelerator science and technology are
multifaceted and challenging, and offer rich rewards for particle
physics.  One line of attack consists in refining known technologies
to accelerate and collide the traditional projectiles---electrons,
protons, and their antiparticles---pushing the frontiers of energy,
sensitivity, and precise control.  The new instruments might include
brighter proton sources; very-high-luminosity $e^{+}e^{-}$
``factories'' for $B$, $\tau$ / charm, $\phi$, \ldots; cost-effective
hadron colliders beyond the LHC at CERN; and $e^{+}e^{-}$ linear
colliders.

A second approach entails the development of exotic acceleration
technologies for standard particles: electrons, protons, and their 
antiparticles.  We don't yet know what
instruments might result from research into new acceleration methods,
but it is easy to imagine dramatic new possibilities for particle
physics, condensed matter physics, applied science, medical
diagnostics and therapies, and manufacturing, as well as a multitude of
security applications.

A third path involves the exploration of exotic particles for
accelerators and colliders to expand the experimenter's 
armamentarium.  Muon storage rings for neutrino factories,
$\mu^{+}\mu^{-}$ colliders and $\gamma\gamma$ colliders are all under 
active investigation, and each of these would bring remarkable new 
possibilities for experiment.

Finally, let us note the continuing importance of enabling 
technologies: developing or domesticating new materials, 
new construction methods, new instrumentation, and new active 
controls.  

To a very great extent, the progress of particle physics has been
paced by progress in accelerator science and technology.  A renewed
commitment to accelerator research and development~\cite{DPBdoc} will
ensure a vigorous intellectual life for accelerator science and lead
to important new tools for particle physics and beyond.

\section{CONCLUDING REMARKS}
In the midst of a revolution in our conception of Nature, we confront
many fundamental questions about our world of diversity and change. 
Are the quarks and leptons elementary or composite?  What are the
symmetries of Nature, and how are they hidden from us?  Will we find
new forms of matter, like the superpartners suggested by
supersymmetry?  Will we find additional fundamental forces?  What
makes an electron an electron and a top quark a top quark?  What is
the dimensionality of spacetime,  what is its shape?

These are themselves great questions and, in the usual way of science,
answering them can lead us toward the answers to yet broader and more
cosmic questions.  As we contemplate the far-reaching understanding we can
hope to create together, it is inspiring to remember the words Michael
Faraday recorded in his
\textit{Research Notes} of 19th March 1849:
\begin{quote}
    Nothing is too wonderful to be true,\\ if it be consistent with the laws 
    of nature \ldots \\
    Experiment is the best test \ldots
\end{quote}

\section*{ACKNOWLEDGMENT}
Fermilab is operated by Universities Research Association Inc.\ under
Contract No.\ DE-AC02-76CH03000 with the United States Department of
Energy.

\end{document}